\documentclass{ringb99}
\usepackage{graphics}
\begin{document}
\renewcommand{\textfraction}{0}  
\renewcommand{\topfraction}{1}
\renewcommand{\bottomfraction}{1}

\title{SPH simulations of magnetic fields in galaxy clusters}
\author{Klaus Dolag\inst{1} \and Matthias Bartelmann\inst{1} \and
  Harald Lesch\inst{2}}
\institute{Max-Planck-Institut f\"ur Astrophysik, P.O.~Box 1523, 
  D--85740 Garching, Germany \and Universit\"ats-Sternwarte 
  M\"unchen, Scheinerstr.~1, D-81679 M\"unchen, Germany}
\maketitle

\begin{abstract}
We perform cosmological, hydrodynamic simulations of magnetic fields
in galaxy clusters. The computational code combines the
special-purpose hardware Grape for calculating gravitational
interaction, and smooth-particle hydrodynamics for the gas
component. We employ the usual MHD equations for the evolution of the
magnetic field in an ideally conducting plasma. As a first
application, we focus on the question what kind of initial magnetic
fields yield final field configurations within clusters which are
compatible with Faraday-rotation measurements. Our main results can be
summarised as follows: (i) Initial magnetic field strengths are
amplified by approximately three orders of magnitude in cluster cores,
one order of magnitude above the expectation from spherical
collapse. (ii) Vastly different initial field configurations
(homogeneous or chaotic) yield results that cannot significantly be
distinguished. (iii) Micro-Gauss fields and Faraday-rotation
observations are well reproduced in our simulations starting from
initial magnetic fields of $\sim10^{-9}\,{\rm G}$ strength. Our
results show that (i) shear flows in clusters are crucial for
amplifying magnetic fields beyond simple compression, (ii) final field
configurations in clusters are dominated by the cluster collapse
rather than by the initial configuration, and (iii) initial magnetic
fields of order $10^{-9}\,{\rm G}$ are required to match
Faraday-rotation observations in real clusters.
\end{abstract}

\section{\label{sec:1} Introduction}

Magnetic fields in galaxy clusters are inferred from observations of
diffuse radio haloes (Kronberg 1994) and Faraday rotation (Vallee,
MacLeod \& Broten 1986, 1987).

It is therefore evident that clusters of galaxies are pervaded by
magnetic fields of $\sim\mu{\rm G}$ strength.


Various models have been proposed for the origin of cluster magnetic
fields in individual galaxies (Rephaeli 1988; Ruzmaikin, Sokolov \&
Shukurov 1989). The basic argument behind such models is that the
metal abundances in the intra-cluster plasma are comparable to solar
values. The plasma therefore must have been substantially enriched by
galactic winds which could at the same time have blown the galactic
magnetic fields into intra-cluster space.

We here address the question whether seed magnetic fields of
speculative origin, magnified in the collapse of cosmic material into
galaxy clusters, can reproduce a number of observations, particularly
Faraday-rotation measurements. Specifically, we ask what initial
conditions we must require for the seed fields in order to
reproduce the statistics of rotation-measure observations. A
compilation of available observations of this kind was published by
Kim, Kronberg \& Tribble (1991).

\section{GrapeMSPH}
\begin{figure*}
 \begin{center}
 \resizebox{0.8\hsize}{!}{\includegraphics{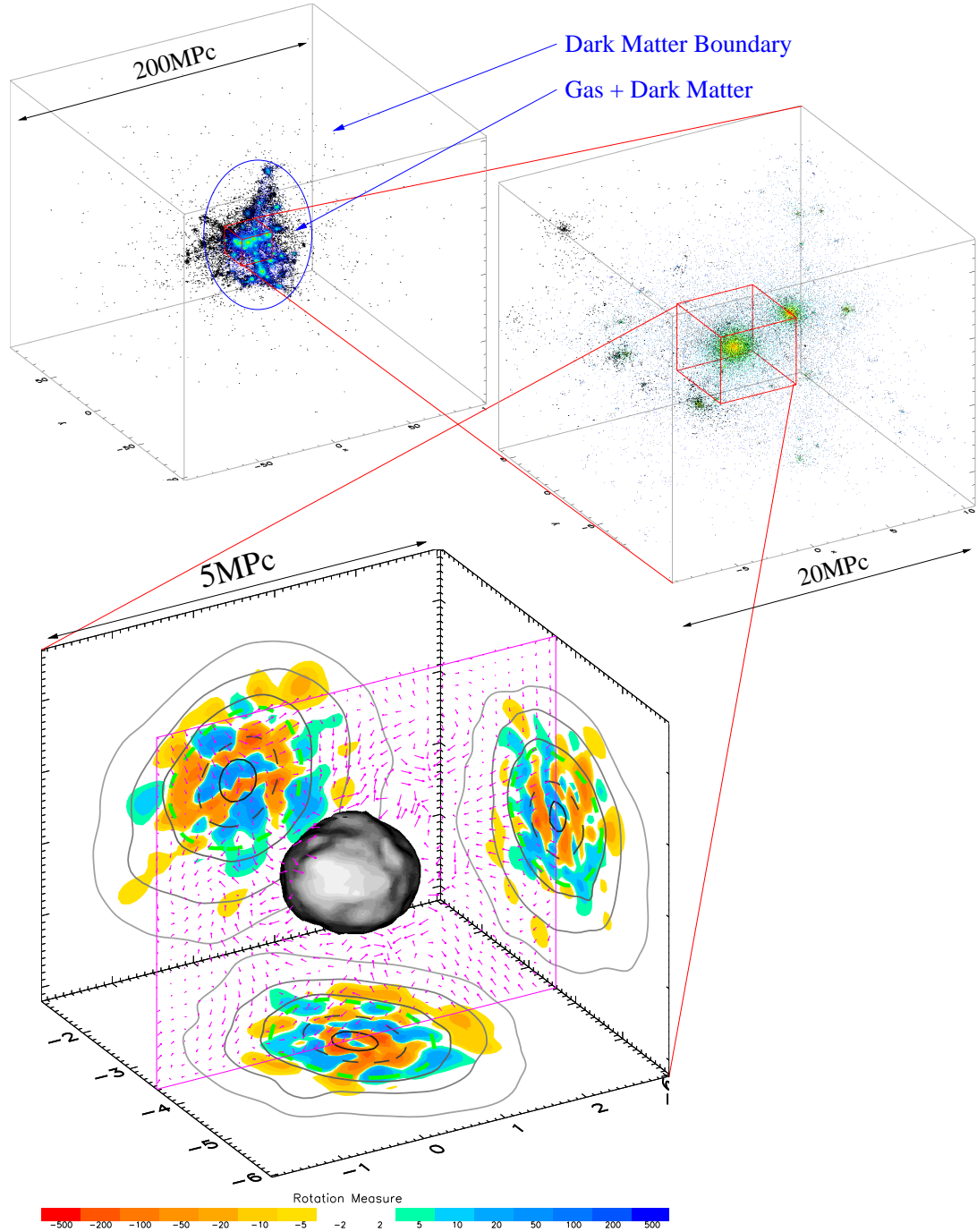}}
 \end{center}
\caption[]{Our simulations work with three classes of particles. In a central
region, we have $\sim50,000$ collision-less dark-matter particles with
mass $3.2\times10^{11}\,M_\odot$, mixed with an equal number of gas
particles whose mass is twenty times smaller. These gas particles
carry the magnetic field. The
central region is surrounded by $\sim20,000$ collision-less boundary
particles whose mass increases outward to mimic the tidal forces of
the neighbouring large scale structure. The figure shows the structure
of this kind of simulation at redshift $z=0$ from the whole simulated 
volume down to the cluster. 
The lower box in this figure shows one simulated cluster in a
three-dimensional box and also appears as color print.
The Faraday-rotation measures produced by the
cluster in the three independent spatial directions are projected onto
the box sides and encoded by the color-scale as indicated below the
box. The gray solid curves are projected density contours, whereas
the dashed line marks half the central density. 
The green dashed curve encompasses the region emitting
90\% of the projected X-ray luminosity. The shaded object in the
center is the isodensity surface at hundred times the critical density. 
In addition, magnetic field vectors are plotted in the slice marked
by the purple rectangle. Coordinates are physical
coordinates in Mpc.}
\label{fig:all}
\end{figure*}

We start with the GrapeSPH code developed and kindly provided by
Matthias Steinmetz (Steinmetz 1996). It simultaneously computes with 
a multiple time-step scheme the behaviour
of two matter components, a dissipation-free dark matter component
interacting only through gravity, and a dissipational, gaseous
component. The gravitational interaction is evaluated on the Grape
board, while the hydrodynamics is calculated by the CPU of the host
work-station in the smooth-particle approach (SPH, Lucy 1977; Monaghan
1985). We supplement the
code with the magneto-hydrodynamic equations to follow the evolution
of an initial magnetic field caused by the flow of the gaseous matter
component. Details on GrapeMSPH and on the results
can be found in Dolag, Bartelmann \& Lesch (1999).
Fig.~\ref{fig:all} shows the appearance of a typical simulation
at the end of a calculation.
 
\section{Initial conditions}
We need two types of initial conditions, namely
(i) the cosmological parameters and initial density perturbations, and
(ii) the properties of the primordial magnetic field. 
For the purposes of the present study, we set up cosmological initial
conditions in an Einstein-de Sitter universe ($\Omega^0_{\rm m}=1$,
$\Omega^0_{\rm \Lambda}=0$) with
a Hubble constant of $H_0=50\,{\rm km\,s^{-1}\,Mpc^{-1}}$. We
initialise density fluctuations according to a COBE-normalised CDM
power spectrum. The origin of the observed cluster magnetic fields is 
uncertain. We therefore use two extreme set-ups for the
initial magnetic field, namely either a chaotic or a completely
homogeneous magnetic field at high redshift ($z=15$). 
For a fair comparison we fixed the average magnetic
field energy density to the same value in both cases.

\section{Results}
The magnetic field in our simulated clusters is dynamically
unimportant even in the densest regions, i.e.~the cluster cores. Since
we ignore cooling, this result may change close to cluster centres
where cooling can become efficient and cooling flows can form.

Magnetic flux conservation in an ideally conducting plasma leads to an
enhancement of the magnetic
field during spherically symmetric cluster collapse proportional to
$\rho^{2/3}$, where $\rho$ is the gas density. 
An initial intracluster field of order $10^{-9}\,{\rm G}$
is therefore expected to be magnified only up to  $10^{-7}\,{\rm G}$.
Our simulations, however, demonstrate that such seed fields can be
magnified up to $10^{-6}\,{\rm G}$ 
in final stages of cluster evolution. {\bf Shear flows therefore
stretch and tangle the
magnetic field, leading to an extra amount of amplification during the
cluster formation process.}

\subsection{Different initial field set-ups}

\begin{figure}
 \begin{center}
 \resizebox{0.7\hsize}{!}{\includegraphics{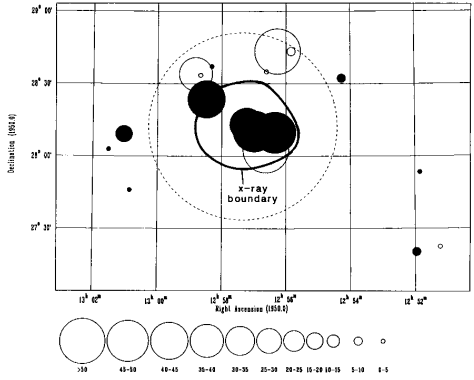}}
 \end{center}
\caption[]{Values and positions of rotation measures observed in 
  the Coma cluster (reproduced from Kim et al.~1990).
}
\label{fig:coma}
\end{figure}

In order to evaluate what the two different kinds of initial
magnetic-field set-ups imply for the observations of rotation
measures (Fig.~\ref{fig:coma}), we compute rotation-measure maps 
from the cluster simulations (lower part of Fig.~\ref{fig:all}) 
and compare them statistically. We use two methods for
that. First is the usual Kolmogorov-Smirnov test, which evaluates the
probability with which two sets of data can have been drawn from the
same parent distribution. Second is an excursion-set approach, in
which we compute the fraction ${\cal F}$ of the total cluster surface
covered by RM values exceeding a certain threshold. For the cluster
surface area, we take the area of the region emitting 90\% of the
X-ray luminosity.

\begin{figure}
 \begin{center}
 \resizebox{0.8\hsize}{!}{\includegraphics{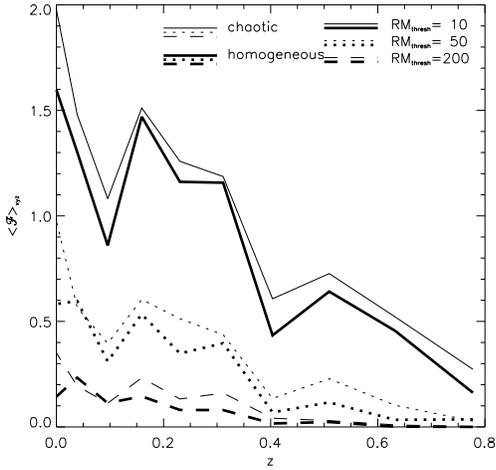}}
 \end{center}
\caption[]{This figure shows the fraction ${\cal F}$ of the cluster area
  which is covered by regions in which the absolute value of the
  rotation measure exceeds a certain threshold for homogeneous (thick
  line) and chaotic (thin line) initial magnetic fields. The cluster area is
  taken to be the area of the region emitting 90\% of the X-ray
  luminosity. The results are averaged over the three independent
  spatial directions. Line types distinguish different RM thresholds,
  as indicated in the figure.}
\label{fig:5}
\end{figure}

Fig.~\ref{fig:5} shows how the RM distributions evolve in
clusters in which the initial magnetic field was set up either
homogeneously or chaotically. We use the excursion-set approach here,
i.e.~we plot the fraction ${\cal F}$ of the cluster area that is
covered by regions in which the RM exceeds a certain threshold.

Quite generally, this fractional area for fixed RM threshold increases
during the process of cluster formation. {\bf As Fig.~\ref{fig:5} shows,
the difference between initially homogeneous (heavy lines) and chaotic
(thin lines) magnetic fields is negligible.} 

\subsection{Individual cluster}

\begin{figure}
 \begin{center}
 \resizebox{0.8\hsize}{!}{\includegraphics{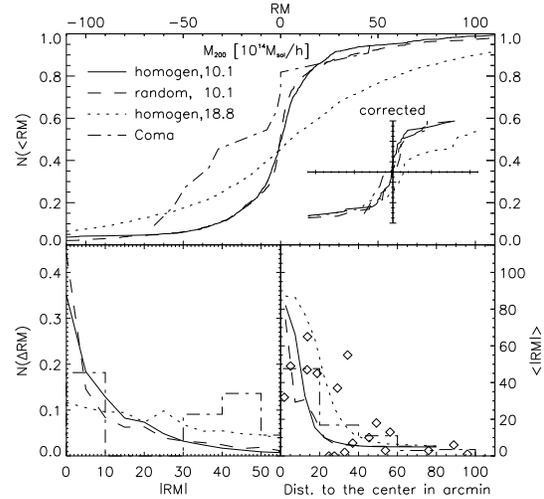}}
 \end{center}
\caption[]{This figure shows different visualizations of the RM statistics for a
simulated cluster. Shown are a Cluster with homogeneous (solid line) and chaotic
(dashed line) initial magnetic field, an approximately twice as
massive cluster (dotted line), and observations in the
Coma cluster (dashed-dotted line or points).
The top panel shows cumulative distributions of rotation
measures. A Kolmogorov-Smirnov test reveals a thirty percent chance for the 
observations to be drawn from the simulated distributions for the
smaller and only one per cent for the more massive cluster, whereas the
homogeneous and chaotic initial conditions cannot be distinguished.
The additional plot in the upper panel shows the final comparison with
the Coma measurements, applying some corrections for intrinsic Faraday rotation
and spatial distribution of the sources.
The lower left panel shows the distribution of the RM measurements
inside the X-Ray boundary. The more massive cluster has relatively
more high and fewer low RM values than the less massive one.
The lower right panel shows the mean of the absolute value for the RM
measurements as function of the distance to the cluster
center. Towards its center, the more massive cluster has higher RM
values than Coma, while the less massive one almost reproduces
the observations. In the outer parts, both clusters produce somewhat
too small RMs.}
\label{fig:comp}
\end{figure}

We can now compare the simulated RM statistics with measurements in
individual clusters like Coma, for which a fair number
of measurements is available, as done in Fig.~\ref{fig:comp}.
The distribution of the RM values as a
function of distance to the cluster centre are shown for the Coma
cluster as points in the lower right panel.
The cumulative distribution of these measurements (dash-dotted curve)
is shown in
the upper panel. This distribution can now be compared to the RM
values obtained from the simulations (other lines in
all panels). We choosed random beam
positions under the constraint that their radial distribution match
the observed one. In this example, the Kolmogorov-Smirnov likelihood
for 30 simulated beams to match the ten measurements in Coma falls
between 30\% and 50\% for the less massive cluster (solid and dashed
curves), compared to a few per cent for the more massive one (dotted
curve). {\bf Since the less massive cluster has approximately the
estimated mass of Coma, this demonstrates that our simulations
well reproduce the RM measurements in Coma.} 

\subsection{A sample of clusters}

\begin{figure}
 \begin{center}
 \resizebox{0.8\hsize}{!}{\includegraphics{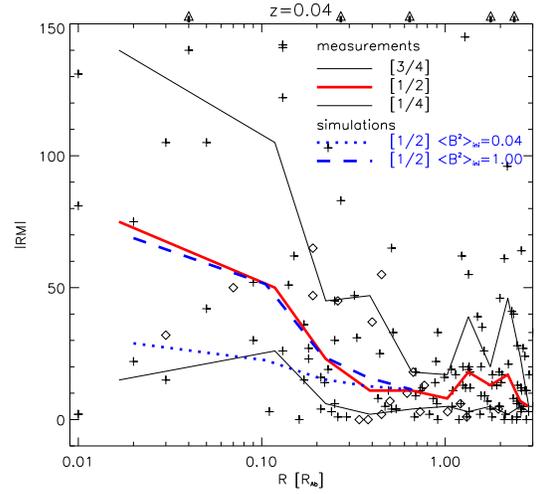}}
 \end{center}
\caption[]{Comparison of simulated results with a sample of measurements
  in Abell clusters. The absolute values of Faraday rotation
  measurements (Kim et al.~1991) vs.~radius in units of the Abell
  radius are shown. Obviously, the dispersion increases towards the
  cluster center. The solid curves mark the median and the 25- and
  75-percentiles of the measurements, and the dashed and dotted curves
  are the medians obtained from simulated cluster samples starting
  with high and low magnetic fields as marked in the plot,
  respectively. The radial bins are chosen such as to
  contain 15 data points each. The scatter in the observations is
  large, but the simulations with the stronger initial magnetic field
  seem to match the observations better.}
\label{fig:8}
\end{figure}

Unfortunately, in most clusters the number of radio sources suitable
for measurements of Faraday rotation is only one or two. We therefore
also compare a sample of simulated clusters with a compilation of RM
observations in different clusters. Fig.~\ref{fig:8} shows the
values published by Kim et al.~(1991). We take the absolute RM value
and plot it against the distance to the nearest Abell cluster in units
of the estimated Abell radius (crosses). 

The increase of the signal towards the cluster centres is evident. In
order not to be dominated by a few outliers, which in some cases even
fall outside the plotted region and are marked by the arrows at the
top of the plot, we did not calculate mean and variance, but rather
the median and the $25$- and $75$-percentiles of the RM distribution
(solid curves). 

The median of synthetic radial rotation-measure distributions obtained
from our simulations for different initial magnetic field strengths is
shown by the dashed and dotted curves. The initial magnetic field
strengths are $10^{-9}{\rm G}$ and $0.2\times10^{-9}{\rm G}$ 
as indicated in the figure. {\bf An initial field strength of order 
$\approx10^{-9}\,{\rm G}$ at redshift $15$ reasonably reproduces
the observations.}

\end{document}